\documentclass[conference]{IEEEtran}
\usepackage{cite}
\usepackage{amsmath}
\usepackage[cmintegrals]{newtxmath}
\usepackage{bm}
\usepackage{graphicx}
\usepackage{url}
\usepackage{dcolumn}
\usepackage{textcomp}
\usepackage{tikz}
\newcommand\copyrighttext{%
  \footnotesize This work has been submitted to the IEEE for possible publication. Copyright may be transferred without notice, after which this version may no longer be accessible. 2021 IEEE. Personal use of this material is permitted. Permission from IEEE must be obtained for all other uses, in any current or future media, including reprinting/republishing this material for advertising or promotional purposes, creating new collective works, for resale or redistribution to servers or lists, or reuse of any copyrighted component of this work in other works.}
\newcommand\copyrightnotice{%
\begin{tikzpicture}[remember picture,overlay]
\node[anchor=south] at (current page.south) {\fbox{\parbox{\dimexpr\textwidth-\fboxsep-\fboxrule\relax}{\copyrighttext}}};
\end{tikzpicture}%
}
\ifCLASSOPTIONcompsoc
  \usepackage[caption=false,font=normalsize,labelfont=sf,textfont=sf]{subfig}
\else
  \usepackage[caption=false,font=footnotesize]{subfig}
\fi
\DeclareMathOperator*{\maximize}{maximize}
\DeclareMathOperator*{\minimize}{minimize}
\title{Evaluation on Energy Efficiency of UE\\in UL Cell-Free Massive MIMO System\\With Power Control Methods}
\author{\IEEEauthorblockN{Masaaki Ito\IEEEauthorrefmark{1}, Issei Kanno\IEEEauthorrefmark{1}, Takeo Ohseki\IEEEauthorrefmark{1}, Kosuke Yamazaki\IEEEauthorrefmark{1}, Yoji Kishi\IEEEauthorrefmark{1},\\Thomas Choi\IEEEauthorrefmark{2}, and Andreas F. Molisch\IEEEauthorrefmark{2}}
\IEEEauthorblockA{\IEEEauthorrefmark{1}KDDI Research, Inc., Saitama, Japan}
\IEEEauthorblockA{\IEEEauthorrefmark{2}University of Southern California, Los Angeles, USA}}
\begin{document}
\copyrightnotice
\maketitle
\begin{abstract}
Cell-free massive multiple-input multiple-output (CF mMIMO) systems are expected to provide faster and more robust connections to user equipments (UEs) by cooperation of a massive number of distributed access points, and to be one of the key technologies for beyond 5G (B5G). In B5G, energy efficiency (EE) is one of the most important key indicators because various kinds of devices connect to the network and communicate with each other. While previously proposed transmit power control methods in CF mMIMO systems have aimed to maximize spectral efficiency or \textit{total} EE, we evaluate in this paper a different approach for maximizing the \textit{minimum} EE among all UEs. We show that this algorithm can provide the optimum solution in polynomial time, and demonstrate with simulations the improved minimum EE compared to conventional methods.
\end{abstract}
\section{Introduction}
In traditional cellular systems, user equipments (UEs) in a particular area called cell are connected only to the antennas of one base station. In contrast, cell-free massive multiple-input multiple-output (CF mMIMO) removes the concept of cell~\cite{ngo}.\footnote{CF mMIMO is also strongly related to the concepts of network MIMO, CoMP, and C-RAN.} Access points (APs) equipped with a single antenna each are distributed in a coverage area, and all APs cooperate together to improve performance of all UEs. Moreover, all APs are connected to the central processing unit (CPU) via backhaul links, and the CPU executes signal processing. CF mMIMO is one of the prospective important technologies for beyond 5G (B5G), or 6G,\footnote{The terms B5G, which we use henceforth, and 6G are used interchangeably in the literature.} because it eliminates inter-cell interference and makes network design more flexible for various use cases of B5G~\cite{demir}.

In B5G scenarios, network optimization pursues two goals: (i) maximization of the spectral efficiency (SE), i.e., to make the best use of the precious resource ``spectrum''; this is also related to the user-experienced data rates (ii) maximization of the whole-network energy efficiency (EE) and EE of each UE. Optimization of whole-network EE is important for environmental reasons and to minimize electricity expenses of the operators, while the EE efficiency of separate UEs because it determines necessary battery size and/or lifetime of a device~\cite{flagship}.

In a realistic scenario, transmit power control (TPC) for uplink and downlink communications is executed to improve performance by reducing interference. Various TPC methods have been proposed; the most common method in CF mMIMO investigations maximizes the minimum SE  among all the UEs; for convenience, we henceforth refer to it as the max-min SE method~\cite{ngo}. In~\cite{ito}, the authors analyzed semi-distributed CF mMIMO systems, including the impact of the max-min SE method. Other proposed TPC methods focus on maximizing EE~\cite{nguyen,nguyen2,alonzo,tran}. However, those papers target the \textit{total}, i.e., the whole network, EE. If total EE is the focus, some UEs can communicate with high EE and others may suffer from low EE, which causes the battery of those UEs to deplete more quickly. 

As we assume that CF mMIMO is deployed for one of the B5G applications mentioned above, high EE of each UE is one of the key requirements to prolong a battery life. Therefore, in this paper, we evaluate different TPC methods in terms of their ability to maximize the \textit{minimum} EE among all the UEs, and demonstrate what improvements can be obtained by methods specifically designed for this goal, compared to other conventional methods. 

Adopting the maximization of the minimum EE as an optimization criterion has been previously proposed in ~\cite{zappone}. In that paper, the total EE and the minimum EE are expressed as global energy efficiency (GEE) and weighted minimum energy efficiency (WMEE), respectively. Although ~\cite{zappone} presents the optimization problems and algorithms to solve them for GEE and WMEE, it does not present evaluations of the WMEE performance. Furthermore, although \cite{hmila,baidas,su} evaluate the minimum EE, these papers do not target distributed antenna systems. Therefore, the performance of TPC to maximize the minimum EE in CF mMIMO systems, which has different propagation characteristics by the antenna distribution, has not been clarified yet.

This paper thus provides a general investigation of the EE performance, especially for respective UEs connecting to the network, of various TPC methods employing commonly used stochastic channel models. Emphasis is on the uplink, which dominates the EE of the UEs.

Notation is as follows: boldface lowercase and uppercase letters denote column vectors and matrices, respectively. Especially, $\bm{0}_M$ denotes an all-zero vector with length of $M$. The superscripts $\mathopen{}\left(\cdot\right)\mathclose{}^{\text{H}}$ and $\mathopen{}\left(\cdot\right)\mathclose{}^{-1}$ denote the Hermitian and inverse matrices, respectively. The Euclidean norm of a vector is denoted by $\mathopen{}\left\|\cdot\right\|\mathclose{}$. Finally, $z\sim\mathcal{N}_{\mathbb{C}}\mathopen{}\left(0,1\right)\mathclose{}$ stands for a complex Gaussian random variable $z$ with mean 0 and variance 1.
\section{System Model}
\begin{figure}[!t]
\centering
\includegraphics[width=2.5in]{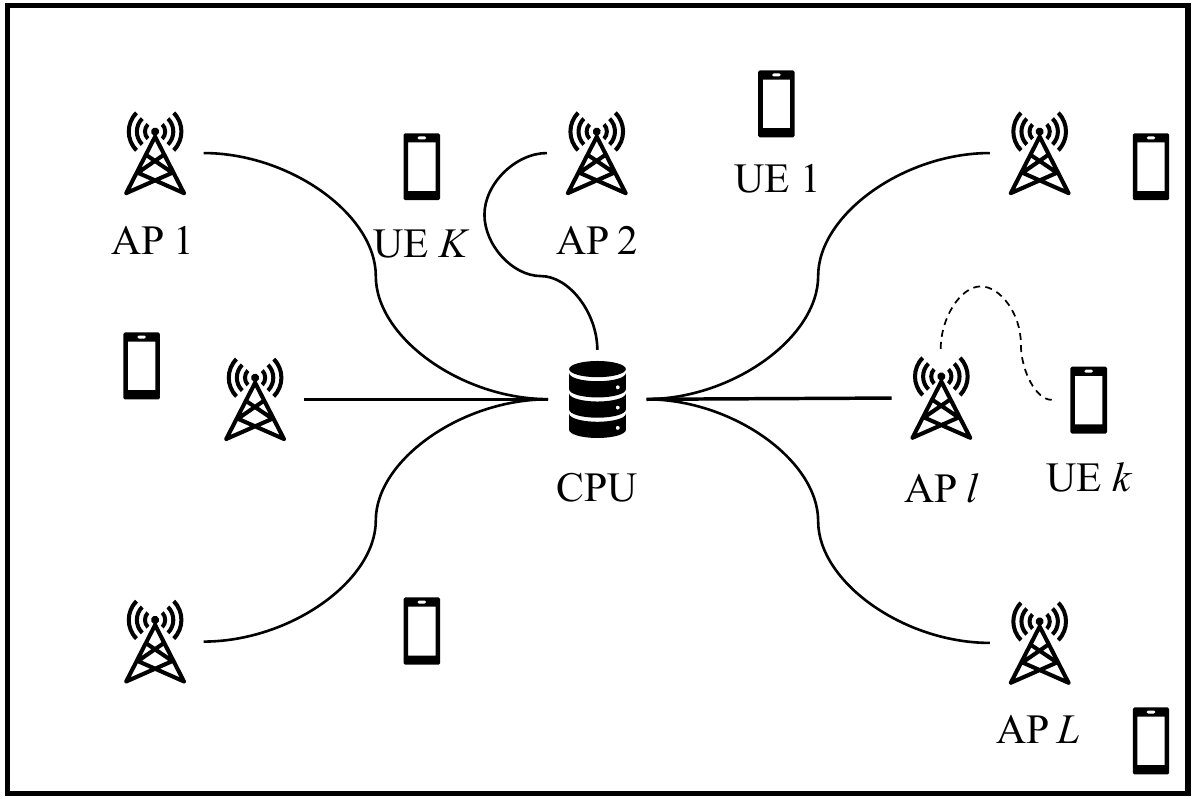}
\caption{System model of CF mMIMO.}
\label{fig:cfmmimo}
\end{figure}

There are $L$ APs deployed in a target area, and each AP has $N$ antennas. Thus, the total number of the network-side antennas is $M$, given as $M=LN$. The network-side antenna index $m$ is defined as $m=\mathopen{}\left(l-1\right)\mathclose{}N+n$ ($1\le l\le L$, $1\le n\le N$). The number of spatial multiplexing UEs is assumed to be $K$, and each UE has a single antenna, illustrated in \figurename~\ref{fig:cfmmimo}.

The channel coefficient $h_{m,k}$ between antenna $m$ and UE $k$ can be written as
\begin{IEEEeqnarray}{rCl}
h_{m,k}&=&\sqrt{\beta_{m,k}}p_{m,k},%
\end{IEEEeqnarray}
where $\beta_{m,k}$ and $p_{m,k}$ are large- and small-scale fading of the corresponding links, respectively. As is common in the literature, we assume flat fading.
\subsection{Uplink System Model}
The received signal at antenna $m$ of symbol $i$ is written as
\begin{IEEEeqnarray}{rCl}
y_m\mathopen{}\left(i\right)\mathclose{}&=&\sqrt{\rho}\sum_{k=1}^K h_{m,k}\sqrt{q_k}s_k\mathopen{}\left(i\right)\mathclose{}+z_m\mathopen{}\left(i\right)\mathclose{},%
\end{IEEEeqnarray}
where $s_k\mathopen{}\left(i\right)\mathclose{}$ is a transmitted symbol of UE $k$ normalized to unit average power, and its transmit power coefficient is $q_k$, i.e., the maximum value is 1. $z_m\mathopen{}\left(i\right)\mathclose{}\sim\mathcal{N}_{\mathbb{C}}\mathopen{}\left(0,1\right)\mathclose{}$ is the normalized noise, and $\rho$ is the transmit SNR, i.e., the ratio of the maximum transmitted signal power divided by the noise power.
\subsection{Channel Estimation}
For channel estimation, $\tau^{\mathopen{}\left(\text{p}\right)\mathclose{}}$ pilot resources are consumed within the coherence interval, and all the UEs transmit pilot signals in the resources. Let $\sqrt{\tau^{\mathopen{}\left(\text{p}\right)\mathclose{}}}\bm{\varphi}_k$ be the $\tau^{\mathopen{}\left(\text{p}\right)\mathclose{}}$-dimensional pilot sequence vector of UE $k$, where $\mathopen{}\left\|\bm{\varphi}_k\right\|\mathclose{}^2=1$, and the corresponding received signal vector can be written as
\begin{IEEEeqnarray}{rCl}
\bm{y}_m^{\mathopen{}\left(\text{p}\right)\mathclose{}}&=&\sqrt{\rho^{\mathopen{}\left(\text{p}\right)\mathclose{}}\tau^{\mathopen{}\left(\text{p}\right)\mathclose{}}}\sum_{k=1}^Kh_{m,k}\bm{\varphi}_k+\bm{z}_m^{\mathopen{}\left(\text{p}\right)\mathclose{}}.%
\end{IEEEeqnarray}

The MMSE channel estimate can be written as
\begin{IEEEeqnarray}{rCl}
\Hat{h}_{m,k}&=&\frac{\sqrt{\rho^{\mathopen{}\left(\text{p}\right)\mathclose{}}\tau^{\mathopen{}\left(\text{p}\right)\mathclose{}}}\beta_{m,k}}{\rho^{\mathopen{}\left(\text{p}\right)\mathclose{}}\tau^{\mathopen{}\left(\text{p}\right)\mathclose{}}\sum_{k'=1}^K\beta_{m,k'}\mathopen{}\left|\bm{\varphi}_k^\text{H}\bm{\varphi}_{k'}\right|\mathclose{}^2+1}\bm{\varphi}_k^{\text{H}}\bm{y}_m^{\mathopen{}\left(\text{p}\right)\mathclose{}}.%
\end{IEEEeqnarray}
\section{Performance Metric}
To analyze the performance of TPC, we evaluate EE.
\subsection{Spectral Efficiency}
For zero-forcing (ZF) reception, the CPU collects received signals of all antennas, and the vectorized received signal can  be written as follows:
\begin{IEEEeqnarray}{rCl}
\bm{y}\mathopen{}\left(i\right)\mathclose{}=\sqrt{\rho}\bm{H}\bm{Q}^{1/2}\bm{s}\mathopen{}\left(i\right)\mathclose{}+\bm{z}\mathopen{}\left(i\right)\mathclose{},%
\end{IEEEeqnarray}
where $\bm{H}$ is the $M\times K$ matrix whose $\mathopen{}\left(m,k\right)\mathclose{}$-element is $h_{m,k}$. $\bm{Q}$ is the $K\times K$ diagonal matrix whose $\mathopen{}\left(k,k\right)\mathclose{}$-element is $q_k$.

For ZF detection, the $K\times M$ weight matrix is formulated as
\begin{IEEEeqnarray}{rCl}
\bm{W}^{\mathopen{}\left(\text{ZF}\right)\mathclose{}}&=&\mathopen{}\left(\Hat{\bm{H}}^{\text{H}}\Hat{\bm{H}}\right)\mathclose{}^{-1}\Hat{\bm{H}}^{\text{H}},%
\end{IEEEeqnarray}
where $\Hat{\bm{H}}$ is the estimate of $\bm{H}$, and the channel estimation error matrix is defined as $\Tilde{\bm{H}}=\bm{H}-\Hat{\bm{H}}$, i.e., $h_{m,k}=\Hat{h}_{m,k}+\tilde{h}_{m,k}$.

The spectral efficiency of UE $k$ with ZF is formulated as
\begin{IEEEeqnarray}{rCl}
S_k^{\mathopen{}\left(\text{ZF}\right)\mathclose{}}&=&\log_2\mathopen{}\left(1+\frac{\rho q_k}{\rho\sum_{k'\ne k}^Kq_{k'}\mathopen{}\left|\bm{w}_k^{\mathopen{}\left(\text{ZF}\right)\mathclose{}\text{H}}\Tilde{\bm{h}}_{k'}\right|\mathclose{}^2+\mathopen{}\left\|\bm{w}_k^{\mathopen{}\left(\text{ZF}\right)\mathclose{}}\right\|\mathclose{}^2}\right)\mathclose{}.\label{eq:SE}%
\end{IEEEeqnarray}
\subsection{Energy Efficiency}
As mentioned in Section I, it is important to evaluate EE because it is one of the key indicators of B5G.

Based on~\cite{bashar}, we define the total power consumption as
\begin{IEEEeqnarray}{rCl}
P_{\text{total}}&=&\Bar{P}\sum_{k=1}^Kq_k+KP_{\text{U}}\IEEEnonumber\\
&&{+}\:L\mathopen{}\left(P_{\text{fix}}^{\mathopen{}\left(\text{AP}\right)\mathclose{}}+P_{\text{bh}}^{\mathopen{}\left(\text{AP}\right)\mathclose{}}\right)\mathclose{}+M\mathopen{}\left(P_{\text{fix}}^{\mathopen{}\left(\text{ant}\right)\mathclose{}}+P_{\text{bh}}^{\mathopen{}\left(\text{ant}\right)\mathclose{}}\right)\mathclose{},\label{eq:ee_rev}%
\end{IEEEeqnarray}
where $\Bar{P}$ is the maximum transmit power, and $P_{\text{U}}$ is the required power to run circuit components at each UE. $P_{\text{fix}}^{\mathopen{}\left\{\mathopen{}\left(\text{AP}\right)\mathclose{},\mathopen{}\left(\text{ant}\right)\mathclose{}\right\}\mathclose{}}$ and $P_{\text{bh}}^{\mathopen{}\left\{\mathopen{}\left(\text{AP}\right)\mathclose{},\mathopen{}\left(\text{ant}\right)\mathclose{}\right\}\mathclose{}}$ are fixed and backhaul power consumption, respectively, relating to each AP (i.e., the baseline energy consumption of an AP that is independent of the number of antenna elements) and antenna.

The total (whole-network) EE of uplink is given by
\begin{IEEEeqnarray}{rCl}
E_{\text{total}}^{\mathopen{}\left(\text{ZF}\right)\mathclose{}}&=&\frac{\text{Bandwidth}\cdot\sum_{k=1}^Kw_k^{\mathopen{}\left(\text{b}\right)\mathclose{}}S_k^{\mathopen{}\left(\text{ZF}\right)\mathclose{}}}{P_{\text{total}}},\label{eq:ee_nw}%
\end{IEEEeqnarray}
where $w_k^{\mathopen{}\left(\text{b}\right)\mathclose{}}$ is the weight for each UE. These weights can be chosen depending on the application and target, e.g., if the goal is to maximize the minimum lifetime of the UE, the weight can be chosen proportional to the remaining battery charge. 

EE of UE $k$ can be obtained as
\begin{IEEEeqnarray}{rCl}
E_k^{\mathopen{}\left(\text{ZF}\right)\mathclose{}}&=&\frac{\text{Bandwidth}\cdot w_k^{\mathopen{}\left(\text{b}\right)\mathclose{}}S_k^{\mathopen{}\left(\text{ZF}\right)\mathclose{}}}{\Bar{P}q_k+P_{\text{U}}}.\label{eq:ee_ue}%
\end{IEEEeqnarray}
\section{Conventional Transmit Power Control Methods}
TPC is commonly used in practical cellular systems, making its inclusion in EE considerations essential. In this section, we introduce three common methods for performance comparison.
\subsection{Max-Power Method}
Max-power, i.e., each UE transmits signals with the maximum allowed power, is not strictly a TPC method, but we apply it to obtain basic performance, and to compare with other TPC methods. When it is applied, the tendency of EE performance is similar to that of SE performance.
\subsection{Max-Min Spectral Efficiency Method~\cite{ngo}}
Max-min SE is one of the most commonly used TPC methods in CF mMIMO, and aims to maximize the minimum SE among all UEs.

The maximization problem can be written as
\begin{IEEEeqnarray}{ll}
\maximize_{\mathopen{}\left\{q_k\right\}\mathclose{}}&\min_{k=1,\dots,K}S_k^{\mathopen{}\left(\text{ZF}\right)\mathclose{}}\label{eq:prob}\\
\text{subject to }&0\le q_k\le1,k=1,\dots,K.\IEEEnonumber%
\end{IEEEeqnarray}
Since the logarithmic function in \eqref{eq:SE} increases monotonically as SINR becomes larger, the problem \eqref{eq:prob} can be reformulated as follows:
\begin{IEEEeqnarray}{ll}
\maximize_{\mathopen{}\left\{q_k\right\}\mathclose{},t}&t\label{eq:prob2}\\
\text{subject to }&t\le\text{SINR}_k,k=1,\dots,K\IEEEnonumber\\
&0\le q_k\le1,k=1,\dots,K.\IEEEnonumber%
\end{IEEEeqnarray}

As proved in~\cite{cumanan}, the problem \eqref{eq:prob2} can be formulated into a standard geometric programming problem, and can be solved by a solver software, e.g., CVX for MATLAB~\cite{cvx,grant}.
\subsection{Max-Total Energy Efficiency Method~\cite{bashar}}
Max-total EE is a common TPC method in CF mMIMO aiming to maximize the total EE of the area.

The maximization problem can be written as
\begin{IEEEeqnarray}{ll}
\maximize_{\mathopen{}\left\{q_k\right\}\mathclose{}}&E_{\text{total}}^{\mathopen{}\left(\text{ZF}\right)\mathclose{}}\label{eq:max_total_prob}\\
\text{subject to }&S_k^{\mathopen{}\left(\text{ZF}\right)\mathclose{}}\ge S_k^{\mathopen{}\left(\text{r}\right)\mathclose{}},k=1,\dots,K\IEEEnonumber\\
&0\le q_k\le 1,k=1,\dots,K,\IEEEnonumber%
\end{IEEEeqnarray}
where $S_k^{\mathopen{}\left(\text{r}\right)\mathclose{}}$ is the required minimum SE for UE $k$ to ensure a certain level of quality of service. The value of $S_k^{\mathopen{}\left(\text{r}\right)\mathclose{}}$ depends on the use cases of each UE. For simplicity, we assume that $S_k^{\mathopen{}\left(\text{r}\right)\mathclose{}}$ is the common value among all UEs, and denote it as $S^{\mathopen{}\left(\text{r}\right)\mathclose{}}$ (omit the index $k$) henceforth.

Using \eqref{eq:ee_rev} and \eqref{eq:ee_nw}, the problem \eqref{eq:max_total_prob} can be reformulated as
\begin{IEEEeqnarray}{ll}
\maximize_{\mathopen{}\left\{q_k\right\}\mathclose{}}&\frac{\text{Bandwidth}\cdot\sum_{k=1}^Kw_k^{\mathopen{}\left(\text{b}\right)\mathclose{}}S_k^{\mathopen{}\left(\text{ZF}\right)\mathclose{}}}{\Bar{P}\upsilon K+KP_{\text{U}}+LP^{\mathopen{}\left(\text{AP}\right)\mathclose{}}+MP^{\mathopen{}\left(\text{ant}\right)\mathclose{}}}\label{eq:max_total_constant_prob}\\
\text{subject to }&S_k^{\mathopen{}\left(\text{ZF}\right)\mathclose{}}\ge S^{\mathopen{}\left(\text{r}\right)\mathclose{}},k=1,\dots,K\IEEEnonumber\\
&0\le q_k\le 1,k=1,\dots,K,\IEEEnonumber\\
&\sum_{k=1}^Kq_k\le\upsilon K,k=1,\dots,K\IEEEnonumber\\
&\upsilon^*\le\upsilon\le 1,\IEEEnonumber%
\end{IEEEeqnarray}
where $P^{\mathopen{}\left(\text{AP}\right)\mathclose{}}=P_{\text{fix}}^{\mathopen{}\left(\text{AP}\right)\mathclose{}}+P_{\text{bh}}^{\mathopen{}\left(\text{AP}\right)\mathclose{}}$, and $P^{\mathopen{}\left(\text{ant}\right)\mathclose{}}=P_{\text{fix}}^{\mathopen{}\left(\text{ant}\right)\mathclose{}}+P_{\text{bh}}^{\mathopen{}\left(\text{ant}\right)\mathclose{}}$. And to make the problem easier to handle, in \eqref{eq:max_total_constant_prob}, $\sum_{k=1}^Kq_k$ in the denominator is replaced with an auxiliary variable $\upsilon$, and $\upsilon^*$ is the slack variable and given by solving the following minimization problem:
\begin{IEEEeqnarray}{ll}
\minimize_{\mathopen{}\left\{q_k\right\}\mathclose{}}&\sum_{k=1}^Kq_k\label{eq:upsilon_prob}\\
\text{subject to }&S_k^{\mathopen{}\left(\text{ZF}\right)\mathclose{}}\ge S^{\mathopen{}\left(\text{r}\right)\mathclose{}},k=1,\dots,K\IEEEnonumber\\
&0\le q_k\le 1,k=1,\dots,K.\IEEEnonumber%
\end{IEEEeqnarray}
Therefore, $\upsilon^*$ is given by
\begin{IEEEeqnarray}{rCl}
\upsilon^*&=&\frac{\sum_{k=1}^Kq_k^+}{K},%
\end{IEEEeqnarray}
where $q_k^+$ is the optimal solution of the problem \eqref{eq:upsilon_prob}.

It is noted that the objective function of the problem \eqref{eq:max_total_constant_prob} increases monotonically when $\upsilon$ is in the range of $\upsilon^*\le\upsilon\le\upsilon^{\text{opt}}$, and decreases monotonically in the range of $\upsilon^{\text{opt}}\le\upsilon\le1$. Therefore, $\upsilon^{\text{opt}}$ can be obtained by using a simple linear search algorithm, e.g., the hill-climbing algorithm~\cite{he}.

With the replacement of $\sum_{k=1}^Kq_k$ by $\upsilon K$, the denominator of the problem \eqref{eq:max_total_constant_prob} becomes a constant. Therefore, the problem \eqref{eq:max_total_constant_prob} can be further reformulated as follows:
\begin{IEEEeqnarray}{ll}
\minimize_{\mathopen{}\left\{q_k,t_k\right\}\mathclose{}}&\prod_{k=1}^K\frac{1}{t_k}\label{eq:max_total_t_prob}\\
\text{subject to }&S_k^{\mathopen{}\left(\text{ZF}\right)\mathclose{}}\ge S^{\mathopen{}\left(\text{r}\right)\mathclose{}},k=1,\dots,K\IEEEnonumber\\
&0\le q_k\le 1,k=1,\dots,K,\IEEEnonumber\\
&t_k\le\text{SINR}_k,k=1,\dots,K,\IEEEnonumber\\
&\sum_{k=1}^Kq_k\le\upsilon K,k=1,\dots,K\IEEEnonumber\\
&\upsilon^*\le\upsilon\le 1,\IEEEnonumber%
\end{IEEEeqnarray}
where $\text{SINR}_k$ is the fractional part of \eqref{eq:SE}.

Then, the problem \eqref{eq:max_total_t_prob} can be solved as follows:
\begin{enumerate}
\item Find out the optimal value of $\upsilon$ to maximize the total EE using a linear search algorithm.
\item Optimize $q_k$ to maximize the total EE.
\end{enumerate}
\section{Max-Min Energy Efficiency Method}
As shown in Section VI, the three methods described above do not maximize the minimum EE, and EE outage, or the lower EE, performance is degraded. Especially in CF mMIMO systems, the channel condition between each AP and UE differs from each other. Thus, applying a TPC method considering each UE's performance is important. Therefore, in this paper, we focus on the max-min EE method to improve the EE outage performance. Inspired by the formulation of the max-min SE method, the optimization problem of the max-min EE method can be written as
\begin{IEEEeqnarray}{ll}
\maximize_{\mathopen{}\left\{q_k\right\}\mathclose{}}&\min_{k=1,\dots,K}E_k^{\mathopen{}\left(\text{ZF}\right)\mathclose{}}\label{eq:mm_ee_prob}\\
\text{subject to }&S_k^{\mathopen{}\left(\text{ZF}\right)\mathclose{}}\ge S_k^{\mathopen{}\left(\text{r}\right)\mathclose{}},k=1,\dots,K\IEEEnonumber\\
&0\le q_k\le 1,k=1,\dots,K.\IEEEnonumber%
\end{IEEEeqnarray}
As with the max-total EE method, for simplicity, we assume that $S_k^{\mathopen{}\left(\text{r}\right)\mathclose{}}$ is the common value among all UEs, and denote it as $S^{\mathopen{}\left(\text{r}\right)\mathclose{}}$ (omit the index $k$) henceforth.

Using \eqref{eq:ee_ue}, the problem \eqref{eq:mm_ee_prob} can be reformulated as follows:
\begin{IEEEeqnarray}{ll}
\maximize_{\mathopen{}\left\{q_k\right\}\mathclose{}}&\min_{k=1,\dots,K}\frac{\text{Bandwidth}\cdot w_k^{\mathopen{}\left(\text{b}\right)\mathclose{}}S_k^{\mathopen{}\left(\text{ZF}\right)\mathclose{}}}{\Bar{P}q_k+P_{\text{U}}}\label{eq:mm_ee_prob_frac}\\
\text{subject to }&S_k^{\mathopen{}\left(\text{ZF}\right)\mathclose{}}\ge S^{\mathopen{}\left(\text{r}\right)\mathclose{}},k=1,\dots,K\IEEEnonumber\\
&0\le q_k\le 1,k=1,\dots,K.\IEEEnonumber%
\end{IEEEeqnarray}
To make the problem easier to handle, replace $q_k$ in the denominator with an auxiliary variable $\nu$:
\begin{IEEEeqnarray}{ll}
\maximize_{\mathopen{}\left\{q_k\right\}\mathclose{},\nu}&\min_{k=1,\dots,K}\frac{\text{Bandwidth}\cdot w_k^{\mathopen{}\left(\text{b}\right)\mathclose{}}S_k^{\mathopen{}\left(\text{ZF}\right)\mathclose{}}}{\Bar{P}\nu+P_{\text{U}}}\label{eq:mm_ee_prob2}\\
\text{subject to }&S_k^{\mathopen{}\left(\text{ZF}\right)\mathclose{}}\ge S^{\mathopen{}\left(\text{r}\right)\mathclose{}},k=1,\dots,K\IEEEnonumber\\
&0\le q_k\le 1,k=1,\dots,K\IEEEnonumber\\
&q_k\le\nu,k=1,\dots,K\IEEEnonumber\\
&\nu^*\le\nu\le 1,\IEEEnonumber%
\end{IEEEeqnarray}
where $\nu^*$ is the slack variable and given as the maximum $q_k$ obtained by solving the following optimization problem:
\begin{IEEEeqnarray}{ll}
\minimize_{\mathopen{}\left\{q_k\right\}\mathclose{}}&\max_{k=1,\dots,K}q_k\\
\text{subject to }&S_k^{\mathopen{}\left(\text{ZF}\right)\mathclose{}}\ge S^{\mathopen{}\left(\text{r}\right)\mathclose{}},k=1,\dots,K\IEEEnonumber\\
&0\le q_k\le 1,k=1,\dots,K.\IEEEnonumber%
\end{IEEEeqnarray}
It can be proved that the optimal solutions of the problems \eqref{eq:mm_ee_prob_frac} and \eqref{eq:mm_ee_prob2} are equal~\cite{he}.

Finally, the problem \eqref{eq:mm_ee_prob2} can be solved with the same steps as the max-total EE method.
\section{Numerical Evaluation}
Table~\ref{table:spec} shows the fundamental parameter specifications of our simulations. The values listed on the table will be applied unless other values are mentioned specifically. APs and UEs are distributed following a uniform distribution, i.e., a binomial point process. In addition, we assume that the pilot signal of every UE is orthogonal with each other, i.e., no pilot contamination occurs. The Rician K-factor is calculated based on the distance between an AP and a UE, and follows the 3GPP technical report~\cite{factor}. The required minimum SE $S^{\mathopen{}\left(\text{r}\right)\mathclose{}}$ will be mentioned with each result. For simplicity, we set weight $w_k^{\mathopen{}\left(\text{b}\right)\mathclose{}}$ to 1 for all UEs.

In this paper, we fix the total number of antennas ($M$). Therefore, the number of antennas on each AP ($N$) is determined based on the number of APs ($L$) to meet $M=LN$.

As we assume that antenna $m$ is equipped to AP $l$, the large-scale fading is given as follows based on~\cite{emil1,emil2}:
\begin{IEEEeqnarray}{rCl}
\beta_{m,k}&=&\underbrace{g_0-10\gamma\log_{10}\mathopen{}\left(\frac{d_{l,k}}{d_0}\right)\mathclose{}}_{\text{Path Loss}}+\underbrace{\frac{\sigma_w^2}{\sqrt{2}}\mathopen{}\left(w_l^{\text{AP}}+w_k^{\text{UE}}\right)\mathclose{}}_{\text{Shadow Fading}},%
\end{IEEEeqnarray}
where $d_{l,k}$ is the distance between AP $l$ and UE $k$. $w_l^{\text{AP}}$ and $w_k^{\text{UE}}$ are normalized shadow fading of AP $l$ and UE $k$, respectively, and $\sigma_w^2$ is the variance. Although shadowing is related to the link, and not separately of the AP and UE, splitting the total link shadowing into two contributions following~\cite{emil1,emil2} is executed to (approximately) consider the shadowing correlation between different UEs and APs, respectively.

The phase of the line-of-sight component can be determined a geometrical consideration. Then, the small-scale fading for non-line-of-sight (NLOS) channels is given as $p_{m,k}^{\mathopen{}\left(\text{NLOS}\right)\mathclose{}}$, where $\bm{p}_{k}^{\mathopen{}\left(\text{NLOS}\right)\mathclose{}}\sim\mathcal{N}_{\mathbb{C}}\mathopen{}\left(\bm{0}_M,\bm{R}_k\right)\mathclose{}$. $\bm{R}_k$ is a diagonal matrix whose diagonal elements consist of the local spatial correlation $R_{l,k}$~\cite{emilbook}, which, in this paper, is assumed to follow from a Gaussian angular power spectrum.

In this paper, the hill-climbing algorithm is applied to optimize $\upsilon$ and $\nu$. The initial value is set as $\upsilon^{\text{init}}=\upsilon^*$ and $\nu^{\text{init}}=\nu^*$. The step size for each iteration is set to 0.1, and $\upsilon$ and $\nu$ approaches 1. If the obtained total EE or the minimum EE is smaller that that of the previous point, the step size will be divided by 3 and the sign will be inverted, i.e., the point will turn back with a smaller step. The iteration will end if the step size becomes smaller than $10^{-4}$.
\begin{table}[!t]
\renewcommand{\arraystretch}{1.3}
\caption{Basic Parameter Specifications}
\label{table:spec}
\centering
\begin{tabular}{l|c}
\hline
Total number of AP antennas ($M$) & 256\\
Number of UEs ($K$) & 8\\
Area & $1\times 1$~km$^2$\\
Number of UE drops & 500\\
Carrier frequency & 3.5~GHz\\
Bandwidth & 20~MHz\\
Noise power & $-92$~dBm\\
Fading & Rician\\
Rician K-factor in dB & $13-0.03\cdot\text{Distance}$\\
Reference distance ($d_0$) & 1~m\\
Median channel gain at $d_0$ ($g_0$) & $-43.3$~dB\\
Path loss exponent ($\gamma$) & 2\\
Azimuth angular standard deviation ($\sigma_{\phi}$) & 20\textdegree\\
Uplink data power ($\Bar{P}$) & 0.2~W\\
UE circuit power consumption ($P_{\text{U}}$) & 0.1~W\\
\hline
\end{tabular}
\end{table}
\subsection{Effect of the Variable $\nu$ on SE and EE With the Max-Min EE Method}
\begin{figure}[!t]
\centering
\subfloat[SE]{\includegraphics[width=0.405\columnwidth]{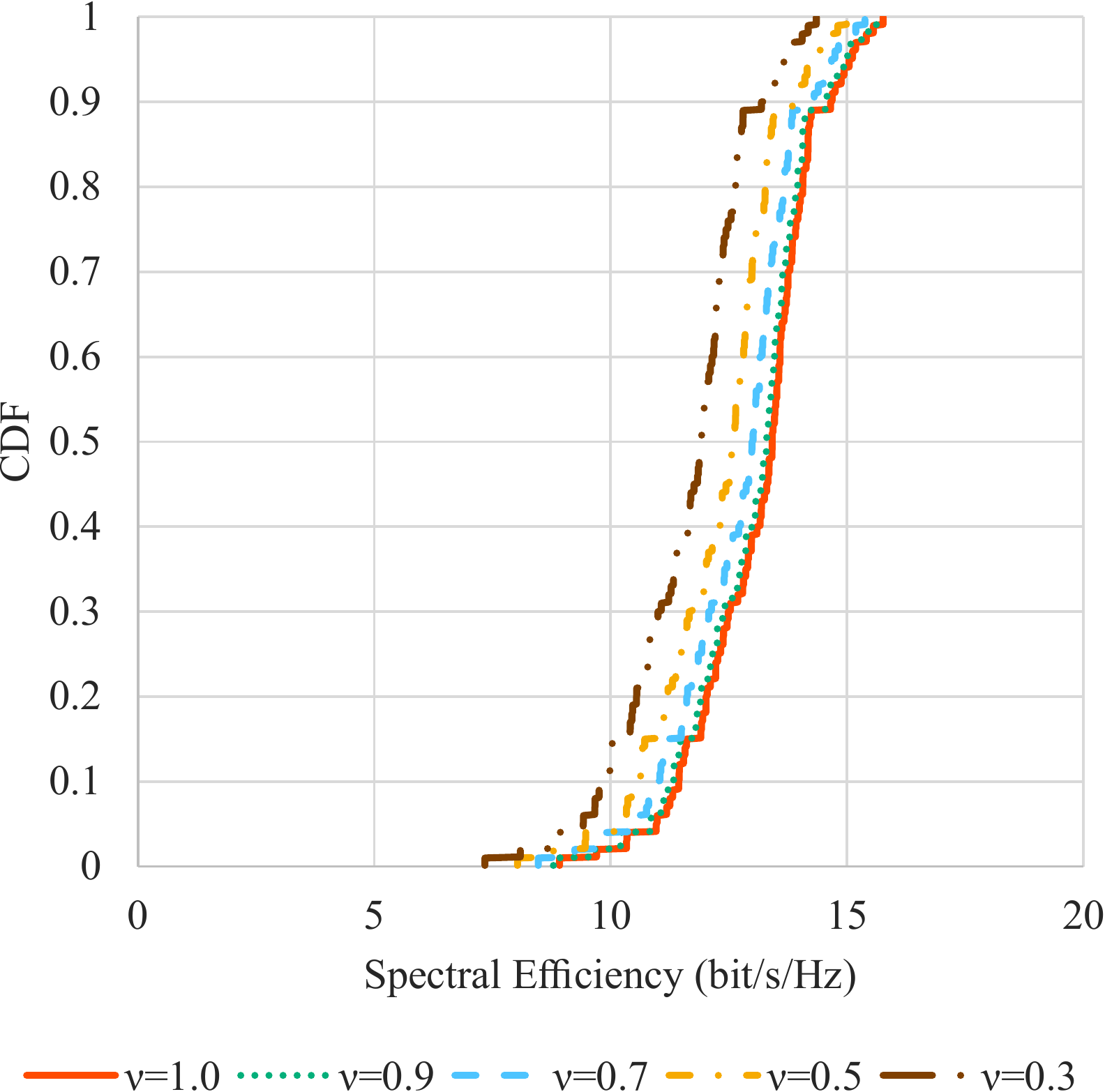}
\label{fig:nu_se}}
\hfil
\subfloat[EE per UE]{\includegraphics[width=0.405\columnwidth]{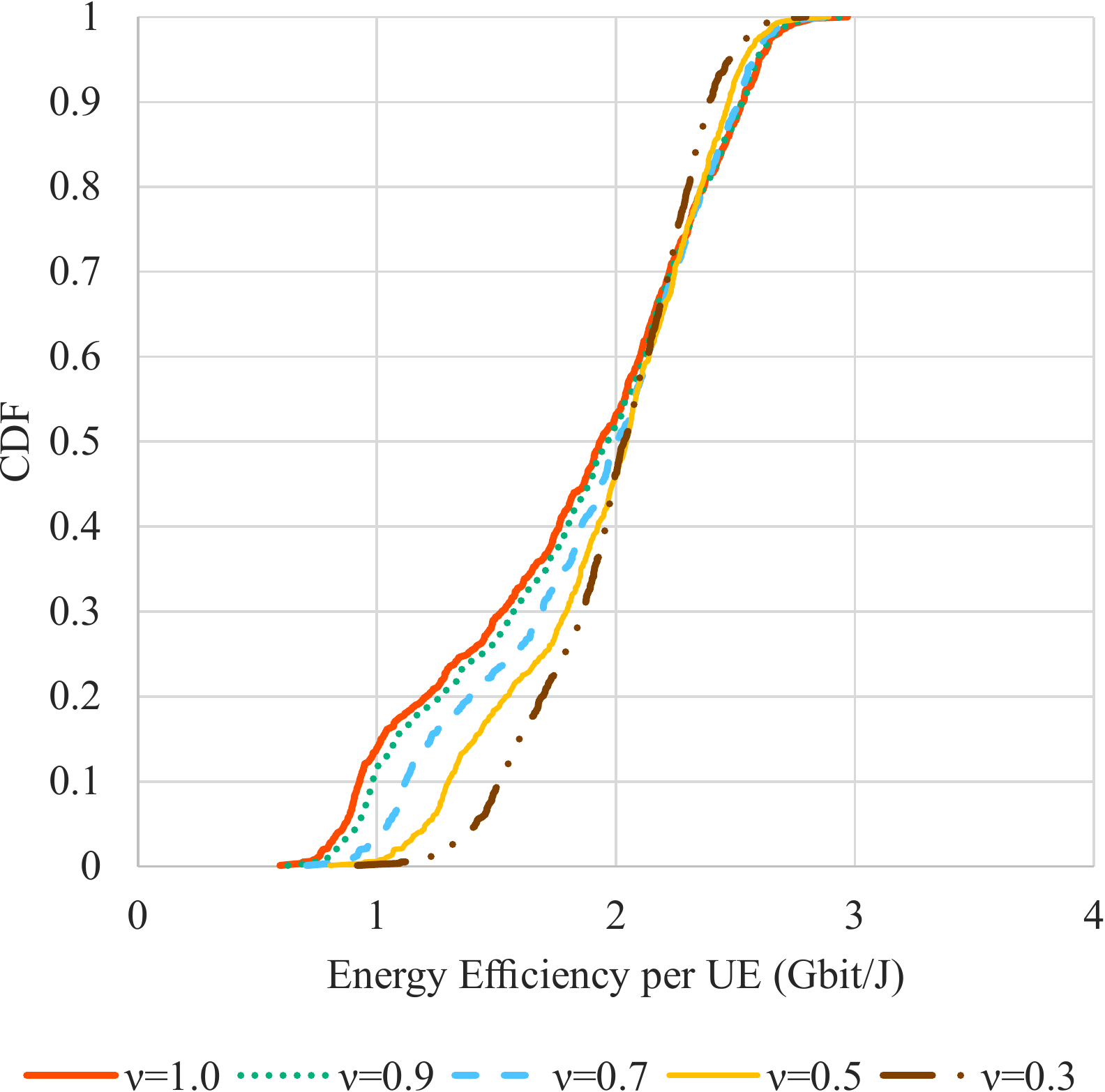}
\label{fig:nu_ee}}
\caption{SE and EE With Various $\nu$.}
\label{fig:various_nu}
\end{figure}

SE and EE with various values of $\nu$ in \eqref{eq:mm_ee_prob2} for $L=256$ and $S^{\mathopen{}\left(\text{r}\right)\mathclose{}}=5$~bit/s/Hz are shown in \figurename~\ref{fig:various_nu}. As can be seen, SE decreases as $\nu$ becomes smaller. This is because $\nu$ is an upper bound value of $q_k$ and the SINR of each UE tends to be lower when the transmit power gets smaller. Even though we set the required SE to 5, the minimum SE is much higher than 5. On the other hand, EE increases as $\nu$ becomes smaller. Since EE is a ratio of SE to power consumption, EE tends to be higher if transmit power decreases more drastically than SE.

Table~\ref{table:min_se_ee} shows the 95\%-likely spectral and energy efficiency for various $\nu$. The SE is 11.0~bit/s/Hz and 9.42~bit/s/Hz for $\nu=1$ and $\nu=0.3$, respectively. That is, the SE performance is degraded by 14\%. In contrast, the EE is 0.865~Gbit/J and 1.41~Gbit/J, respectively. Therefore, the EE performance improves by 63\%. As a result, if communication devices think that EE is more important than SE, $\nu$ should be smaller. This also means that there is room to decrease SE, thus we can improve the minimum EE by finding an optimal value of $\nu$.
\begin{table}[!t]
\renewcommand{\arraystretch}{1.3}
\caption{The 95\%-Likely SE and EE With the Max-Min Method}
\label{table:min_se_ee}
\centering
\begin{tabular}{c||c|c|c|c|c}
\hline
$\nu$ & 1.0 & 0.9 & 0.7 & 0.5 & 0.3\\
\hline\hline
95\%-Likely SE (bit/s/Hz) & 11.0 & 10.8 & 10.5 & 10.1 & 9.42\\
95\%-Likely EE (Gbit/J) & 0.865 & 0.919 & 1.05 & 1.21 & 1.41\\
\hline
\end{tabular}
\end{table}
\subsection{Performance Comparison with Conventional TPC Methods}
\begin{figure}[!t]
\centering
\includegraphics[width=0.405\columnwidth]{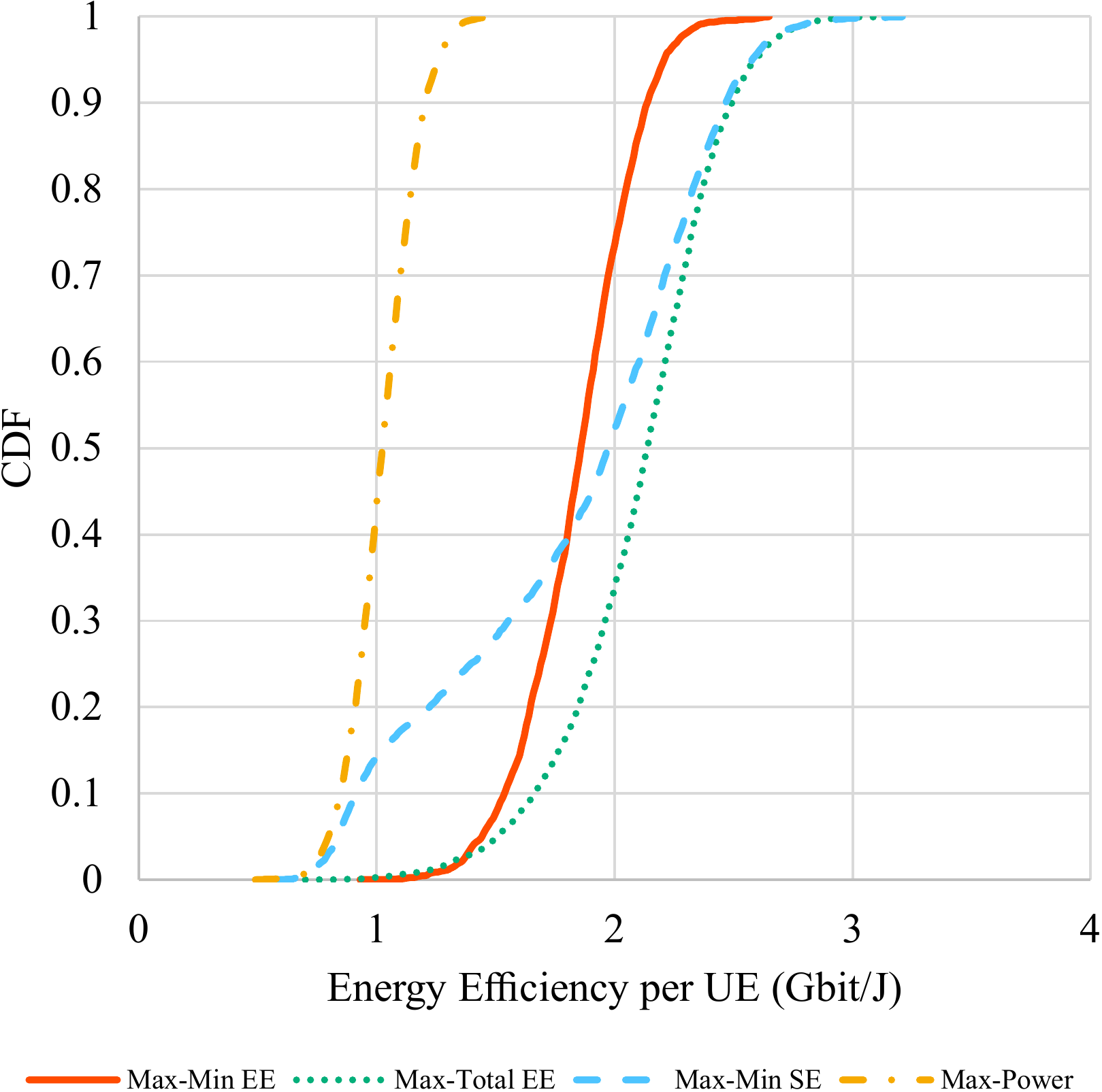}
\caption{Comparison of EE per UE Among Four TPC Methods.}
\label{fig:ee_comparison_5}
\end{figure}

\figurename~\ref{fig:ee_comparison_5} shows EE per UE performance for $L=256$ and $S^{\mathopen{}\left(\text{r}\right)\mathclose{}}=5$~bit/s/Hz of four TPC methods: max-power, max-min SE, max-total EE, and max-min EE. The 95\%-likely EE (Gbit/J) of each TPC method are: 0.795, 0.835, 1.51, and 1.45 for the max-power, the max-min SE, the max-total EE, and the max-min EE, respectively.

As can be seen, the 95\%-likely EE is improved significantly (nearly doubled compared to the Max-Power and the Max-Min SE methods) by applying the max-min EE method. Although the EE of the max-min EE method is lower compared to the max-total EE method, the max-min EE method has a superiority in EE as discussed in the following subsection.
\subsection{Performance Comparison With Various Required SE}
\begin{figure}[!t]
\centering
\subfloat[SE]{\includegraphics[width=0.405\columnwidth]{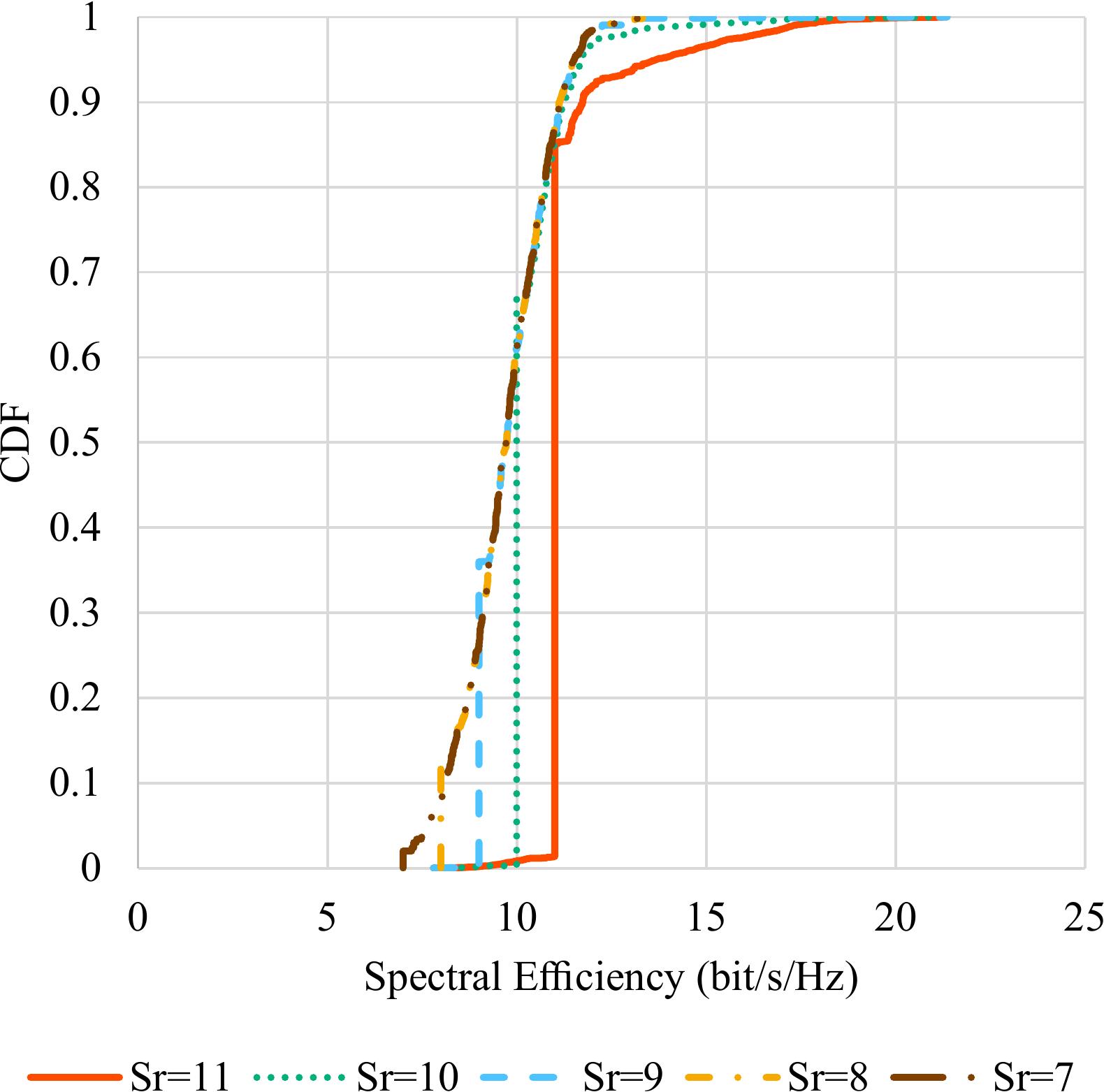}
\label{fig:sr_mm_ee_se}}
\hfil
\subfloat[EE per UE]{\includegraphics[width=0.405\columnwidth]{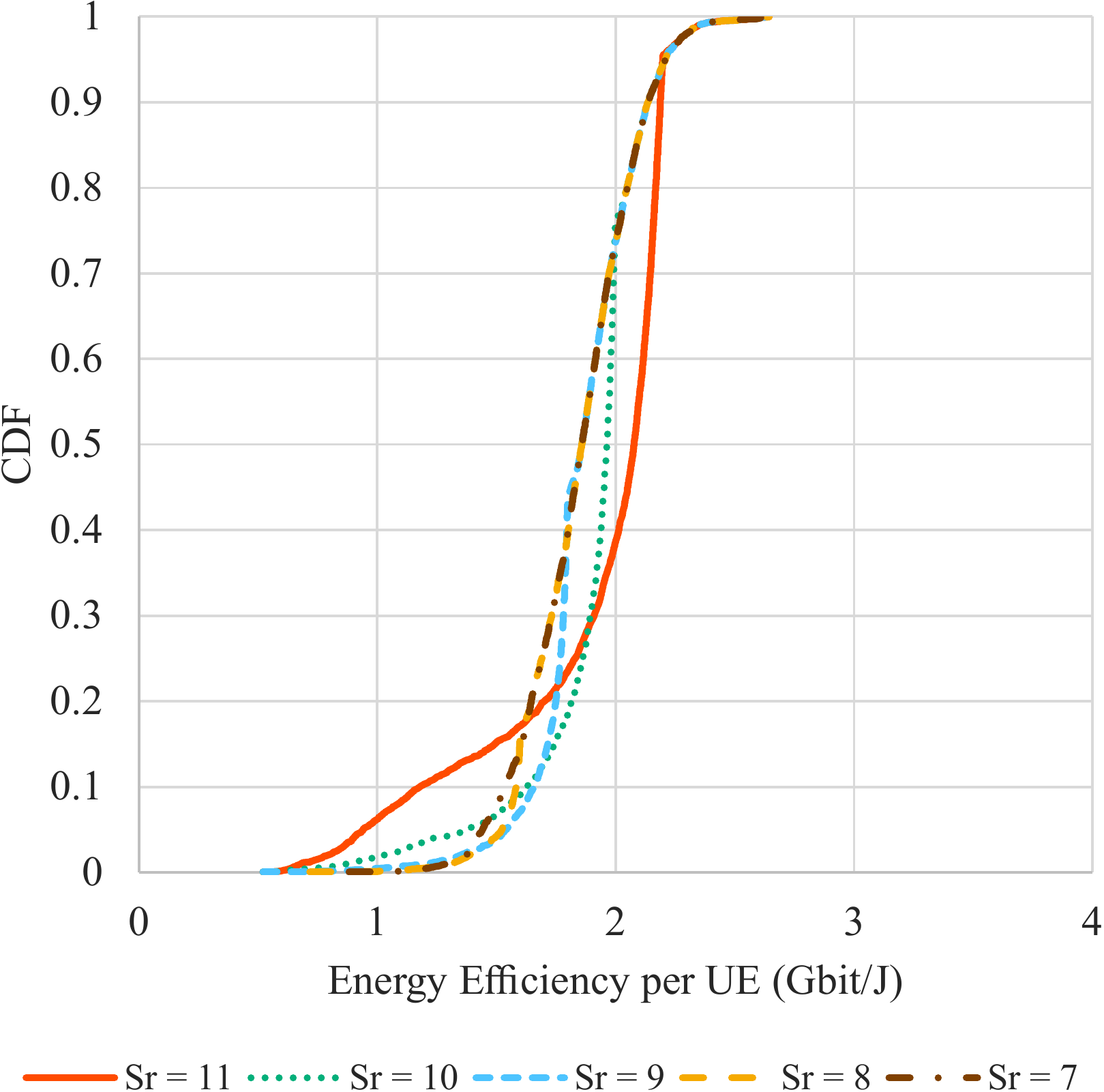}
\label{fig:sr_mm_ee_per_ue}}
\hfil
\subfloat[Total EE]{\includegraphics[width=0.405\columnwidth]{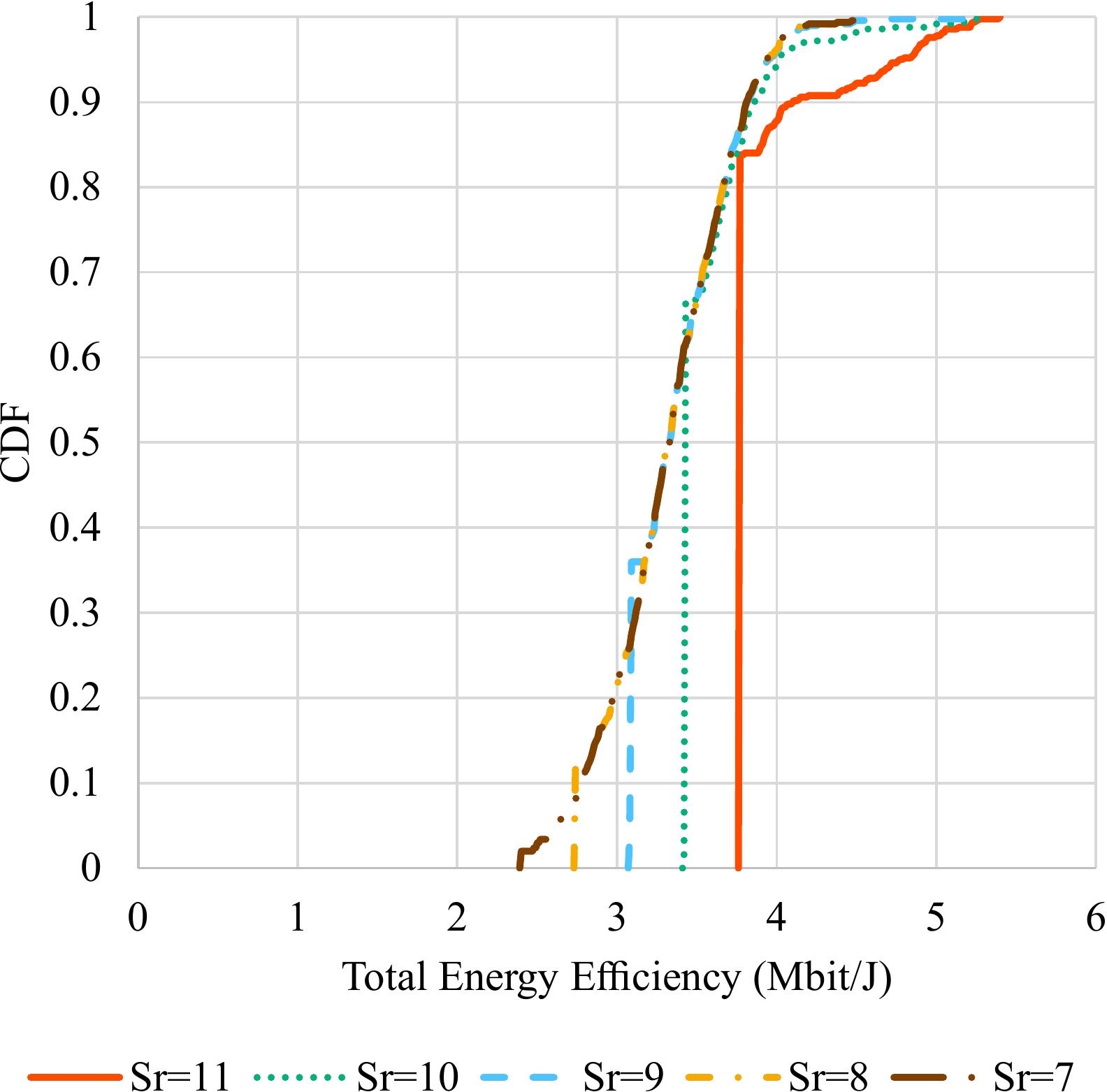}
\label{fig:sr_mm_ee_total}}
\caption{SE and EE Performance of the Max-Min EE Method.}
\label{fig:sr_mm_ee_three}
\end{figure}

\begin{figure}[!t]
\centering
\subfloat[SE]{\includegraphics[width=0.405\columnwidth]{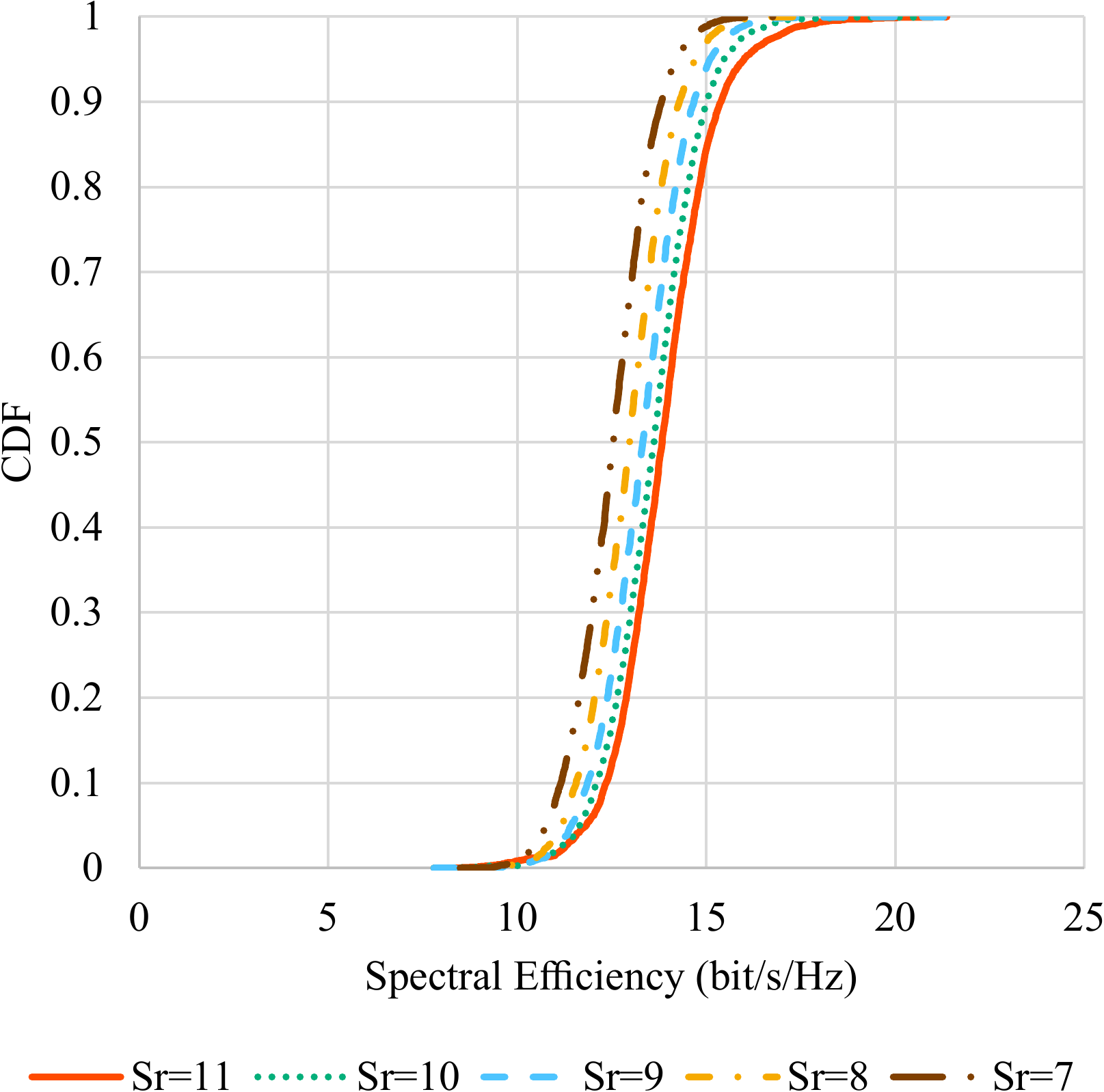}
\label{fig:sr_mt_ee_se}}
\hfil
\subfloat[EE per UE]{\includegraphics[width=0.405\columnwidth]{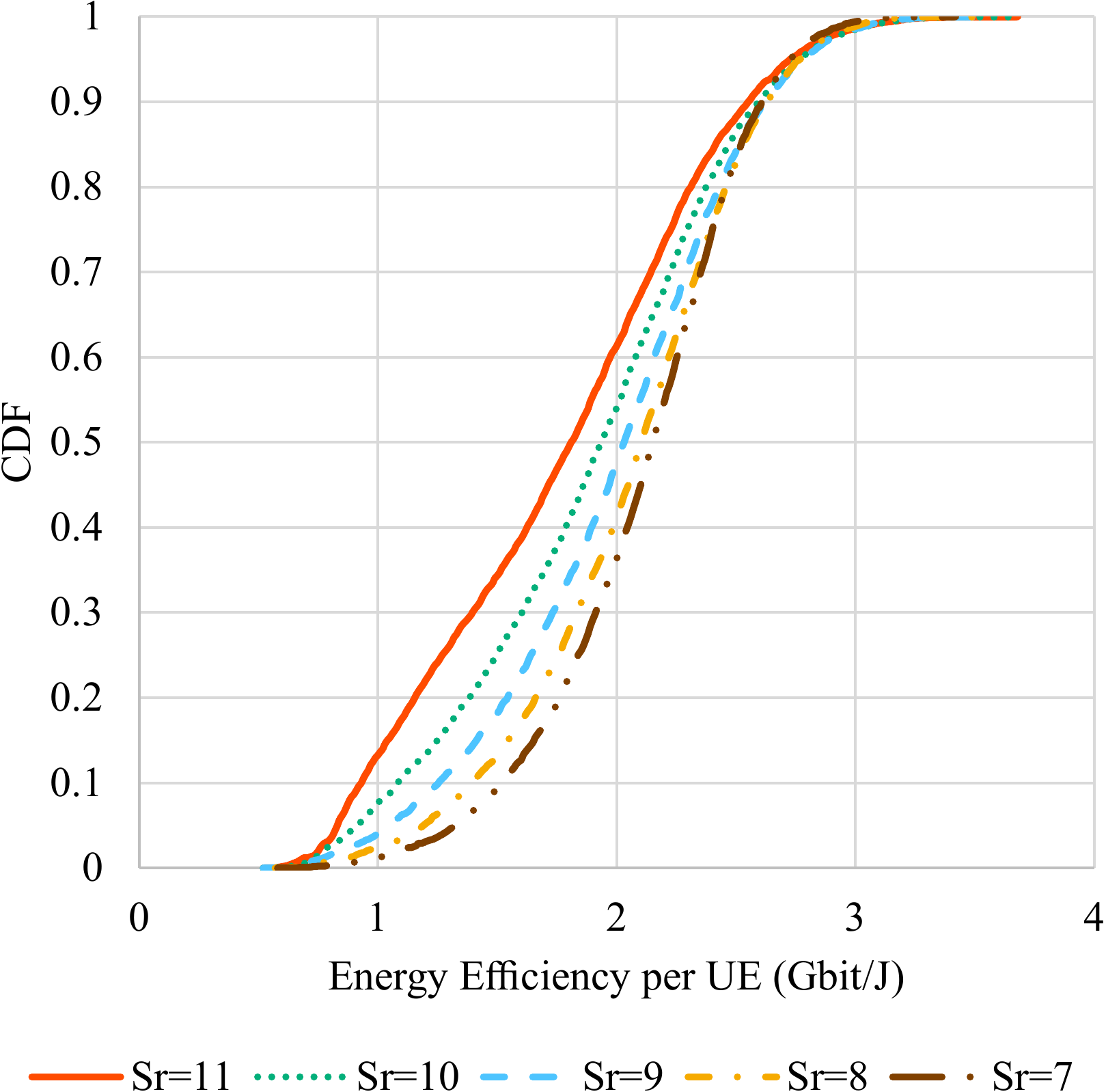}
\label{fig:sr_mt_ee_per_ue}}
\hfil
\subfloat[Total EE]{\includegraphics[width=0.405\columnwidth]{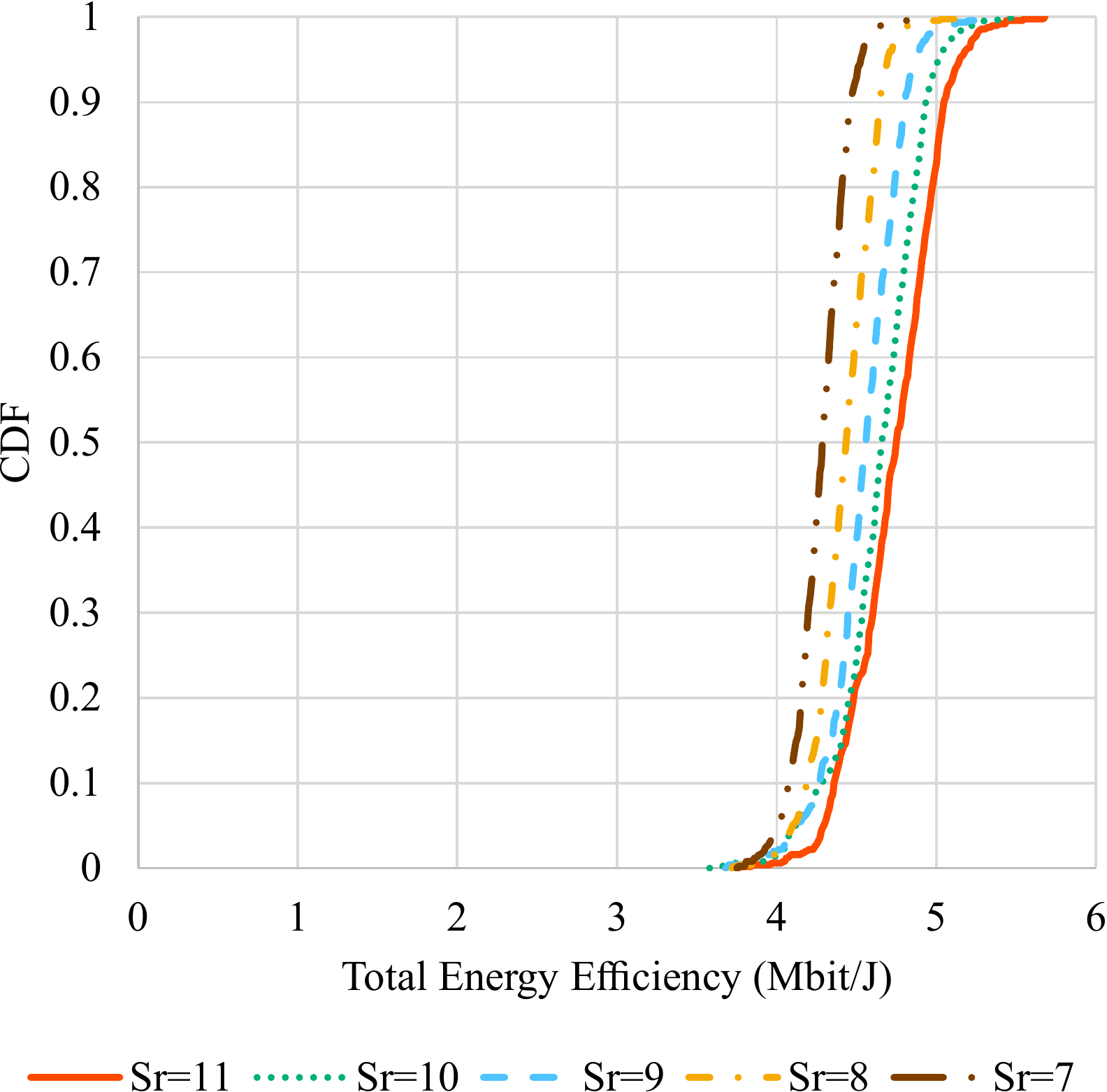}
\label{fig:sr_mt_ee_total}}
\caption{SE and EE Performance of the Max-Total EE Method.}
\label{fig:sr_mt_ee_three}
\end{figure}

\figurename~\ref{fig:sr_mm_ee_three} shows the SE and EE performance of the max-min EE method with different $S^{\mathopen{}\left(\text{r}\right)\mathclose{}}$ for $L=256$, and \figurename~\ref{fig:sr_mt_ee_three} shows the performance of the max-total EE method. In the evaluation of the total EE performance, value of each power consumption in \eqref{eq:ee_rev} is as follows according to~\cite{bashar}: $P_{\text{fix}}^{\mathopen{}\left(\text{AP}\right)\mathclose{}}=0.0825$~W, $P_{\text{fix}}^{\mathopen{}\left(\text{ant}\right)\mathclose{}}=0.743$~W, $P_{\text{bh}}^{\mathopen{}\left(\text{AP}\right)\mathclose{}}=0.1$~W, and $P_{\text{bh}}^{\mathopen{}\left(\text{ant}\right)\mathclose{}}=0.9$~W. The power consumption ratio of an antenna to an AP is 9 to 1. This is based on that an antenna consumes larger power than an AP because of the signal processing effort that is tied to the number of antenna elements. Note that the performance varies according to the ratio.

For the max-min EE, although each curve for different $S^{\mathopen{}\left(\text{r}\right)\mathclose{}}$ begins to rise at a different point of the horizontal axis, they converge to the same curve in the upper part. We can make two observations about the EE of the max-min EE method: (i) although EE and SE can be traded off against each other, EE increases when $S^{\mathopen{}\left(\text{r}\right)\mathclose{}}$ becomes higher. EE can be simplified as the ratio of SE to $q_k$ and the rate at which SE increases is larger than that of $q_k$, and (ii) some UEs cannot achieve $S^{\mathopen{}\left(\text{r}\right)\mathclose{}}$ even though they transmit signals with the maximum power when $S^{\mathopen{}\left(\text{r}\right)\mathclose{}}$ becomes higher. These two effects can be traded off, and the results in \figurename~\ref{fig:sr_mm_ee_three} are determined by the balance of them. For example, the middle part of the curves in \figurename~\ref{fig:sr_mm_ee_three}(b) goes to the right when $S^{\mathopen{}\left(\text{r}\right)\mathclose{}}$ becomes higher because of the observation (i). On the other hand, the lower part of the curves is distorted when $S^{\mathopen{}\left(\text{r}\right)\mathclose{}}$ is high because of the observation (ii).

For the max-total EE, the performance becomes degraded as the required SE increases. Although the upper part of the CDF curves in \figurename~\ref{fig:sr_mt_ee_three}(b) is greater compared to the max-min EE method, the outage performance, in terms of the bottom 5--10\% of the CDF, is worse. The 95\%-likely EE per UE performance of the max-min EE is better than that of the max-total EE, compared to the result in \figurename~\ref{fig:ee_comparison_5} for $S^{\mathopen{}\left(\text{r}\right)\mathclose{}}=5$~bit/s/Hz. In addition, the curves of the max-min EE method in \figurename~\ref{fig:sr_mm_ee_three}(b) is steeper, which means that we can make EE of each UE fairer by applying the max-min EE method.
\section{Conclusion}
In this paper, we evaluated various TPC methods mainly from energy efficiency point of view. Many previous papers on performance evaluation of CF mMIMO systems adopt the total EE as an evaluation metric. However, EE per UE is also one of the key factor because of various use cases in B5G. We compared EE performance of the max-min EE method with other, conventional, methods and showed that the max-min EE method can improve the minimum EE with the constraint of required minimum SE when the APs are distributed over a large area. In addition, the max-min EE method can provide fairer UE performance in terms of EE.

It is clarified that CF mMIMO systems are flexible for various network requirements, and enable us to design not only ``great SE everywhere'' but also ``great EE everywhere'' environments by applying an appropriate TPC method.
\bibliographystyle{IEEEtran}
\bibliography{IEEEabrv,main}
\end{document}